\documentstyle[12pt]{article}
\rm

\def\var{\varepsilon}
\def\f{\frac}
\def\om{\omega}

\def\om{\omega}
\def\inf{\infty}
\def\lam{\lambda}
\begin{document}
\begin{center}
{\large
\bf    Strong-coupling surface polaron in a magnetic field \vspace*{2cm}\\}
       Hua Xiukun\vspace*{1cm}\footnote{xkhua@pub.sz.jsinfo.net}\\
Department of  Physics \\
Suzhou University, Suzhou, 215006,China
\vspace*{2cm}
\end{center}
\begin{center}
{\bf Abstract}
\end{center}
\baselineskip 0.65cm
\hspace*{0.7cm}By applying variational method, the strong electron-surface
optical(SO) phonon interaction and the weak electron-longitudinal optical
(LO) phonon interaction are studied systematically in present paper. The
formula of induced energy $V_{e-so}$, $V_{e-lo}$ in both the ground  
and excited states are given. The numerical results show that as the distance between the
electron and the polar crystal surface increases, the electron-LO phonon
interaction energy increases yet the electron-SO phonon interaction energy
decreases. The numerical results also show that both electron-LO and
electron-SO phonon interaction are enhanced as the magnetic field
strength increases.\\
\\
Key words: Semiconductors, Electron-phonon interactions \\
\newpage
{\bf 1. Introduction} \\
\hspace*{0.5cm} The behavior of polaron has received much attention in
the past decades [1,2].
 Huybrechts [3] proposed a linear combination operator method and investigated
 the property of polaron in strong electron-phonon coupling case. Tokuda [4]
 added another variational parameter to the momentum operator and studied the
 ground-state energy and effective mass of the bulk polaron. Gu [5] discussed
 the the ground-state energy, the first-excited state energy and the effective
 mass of intermediate-coupling polaron in a polaron slab. When external magnetic
 field is applied to the electron-phonon interaction system, the property of
 magnetopolaron attracted many investigators' interests. Whitfield {\it et al} [6] studied
 the effectiveness of adiabatic approximation theory for a polaron in a magnetic
 field and they found that the usual adiabatic theory applied to a polaron in a
 strong magnetic field does not give the right weak coupling limit. Ercelebi [7]
 studied the two-dimensional magnetopolaron in the strong-coupling regime and
 proposed a modified coherent phonon state. Hai, Peeters {\it et al} [8] did research
 into the polaron-cyclotron-resonance spectrum resulting from interface- and
 slab-phonon modes in a GaAs/AlAs quantum well and an experiment was done to
 investigate the resonant magnetopolaron effects due to interface phonons in
 GaAs/AlGaAs multiple quantum well structures by Wang, Nickle {\it et al} [9].
 Recently, the author etc have studied the electron self-energy and effective mass by 
 the powerful Green's function method [10].
\\
 \hspace*{0.5cm} For the bulk polaron, the perturbation theory is applicable
 when $\alpha_l< 1$ and the LLP(Lee-Low-Pines) variational method is effective
 when $\alpha_l <6$. For the surface polaron, Pan [11] showed that when the
 electron-surface optical(SO) phonon coupling constant $\alpha_s >2.5$, the
 strong coupling theory must be applied. Therefor ,in some polar crystals, the
 weak electron-LO phonon coupling and strong electron-SO phonon coupling should
 be considered together. So far, the research work in such a field has been scarce
 \\
 \hspace*{0.5cm} In the present paper, using variational method and taking
 weak electron-LO phonon and strong electron-SO phonon interaction into account,
 we give the ground and excited states energies  of the system, the electron-LO phonon
 interaction induced energy $V_{e-LO}$ and the electron-SO phonon interaction
 induced energy $V_{e-SO}$. Using polar crystal AgBr as an example, numerical
 evaluation are also presented.\\
{\bf 2. Theory}\\
\hspace*{0.5cm}Considering the following system: the semi-infinite space
$z>0$ is occupied by the AgBr crystal, whereas the space $z<0$ is a vacuum.
The external magnetic field is applied along the z direction and an
electron is moving in the polar crystal. {\bf B}=(0,0,B), choosing
the symmetric gauge for the vector potential, i.e. {\bf A}=B/2(-y,x,0), the
Hamiltonian describing an electron coupled to both LO phonons and SO
phonons can be written as($\hbar=m_b$=1, $m_b$ is the band mass of the
electron)
\begin{eqnarray}
& &         H= H_0 +H_\perp +H_{e-ph}=H_\perp+H_{//}\hspace{0.2cm}, \\
& &         H_0=\f{P_{//}^2}{2}+\f{\om_c^2}{8}(x^2+y^2)
         +\f{\om_c}{2}L_z+\sum_k\om_{lo}a_{k}^{\dag}a_k
         +\sum_q\om_{so}b_{q}^{\dag}b_q \hspace{0.2cm},\\
& &         H_\perp=\f{P_{z}^2}{2}+\f{e^{2}(\var_{\inf}-1)}
         {4z\var_{\inf}(\var_{\inf}+1)} \hspace{0.2cm},\\
& &         H_{e-ph}=H_{e-lo}+H_{e-so}\hspace{0.2cm} ,\\
& &         H_{e-lo}= \sum_{k}[V_{k}^{\ast}sin(zk_z)exp(-ik_{//}\cdot\rho)
         a_{k}^{\dag} +h.c.] \hspace{0.2cm},\\
& &         H_{e-so}=\sum_{q}[V_{q}^{\ast}exp(-qz)exp(-iq\cdot\rho)
         b_{q}^{\dag} +h.c.]\hspace{0.2cm}.
\end{eqnarray}
where
\begin{eqnarray}
     & &      V_{k}^{\ast}=i(\f{4\pi e^{2}\om_lo}{\var V})^{\f{1}{2}}\f{1}{k}\hspace{0.1cm},
     \hspace{0.5cm}\f{1}{\var}=\f{1}{\var_{\inf}}-\f{1}{\var_0}\hspace{0.2cm}.   \nonumber \\
       & &   V_{q}^{\ast}=i(\f{\pi e^{2}\om_{so}}{\var^{\ast}qA})^{\f{1}{2}}\hspace{0.1cm},
      \hspace*{0.9cm}       \f{1}{\var^{\ast}}=\f{\var_{0}-1}{\var_{0}+1}
           -\f{\var_{\inf}-1}{\var_{\inf}+1}\hspace{0.2cm}.   \nonumber
\end{eqnarray}
\hspace*{0.5cm} In the above equations,$a_{k}^{\dag}(a_{k})$ is the
creation(annihilation) operator of bulk LO phonons with three-dimensional
wave vector k, $b_{k}^{\dag}(b_{k})$ is the corresponding operator for the
SO phonons with two-dimensional wave vector q.  
$P_{//}=(p_x,p_y)$ and $\rho=(x,y)$
are the electron momentum and position vector in xy plane,respectively,
and $p_z$ is the electron momentum in z direction.
$k_{//}=(k_x, k_y)$ is phonon wave vector in the xy plane. $\om_{lo}$ and
$\om_{so}$ are the frequencies of the bulk LO and SO phonons.$\var_0$ and
$\var_{\inf}$ are the static and high-frequencies dielectric constant of
the crystal respectively. A and V are the surface area and the volume of
the crystal.\\
\hspace*{0.5cm} The polaron system wave function can be separated into
the electron and the phonon parts.
\begin{equation}
         |\psi\rangle=|\phi_{e}\rangle|\phi_{ph}\rangle
\end{equation}
With $|\phi_{ph}\rangle=u_{1}u_{2}|0\rangle$, $|0\rangle$ is the phonon
ground state, and the explicit form of $u_{1}, u_{2}$ will be introduced
in the following.\\
\hspace*{0.5cm}Let us start from the unitary transformation:
\begin{equation}
             u_1=exp(-i(\sum_{k}a_{1}a_{k}^{\dag}a_{k}k_{//}\cdot\rho
             +\sum_{q}a_{2}b_{q}^{\dag}b_{q}q\cdot\rho))
\end{equation}
With $a_{1}=1$ and $a_{2}=0$ corresponding to the weak-coupling limit
case and the strong-coupling limit case respectively. The parts of
polaron Hamiltonian can be transformed into:
\begin{eqnarray}
& & H'_0=u_{1}^{-1}H_{0}u_{1}=\f{1}{2}(P_{//}-\sum_{k}a_{k}^{\dag}
                 a_{k}k_{//})^{2}
             +\f{\om_{c}^2}{8}(x^2+y^2)+\f{\om_c}{2}L_z \nonumber \\
& &\hspace*{1cm} +\f{\om_c}{2}\sum_{k}a_{k}^{\dag}a_{k}
             (yk_{x}-xk_{y})+\sum_{k}\om_{lo}a_{k}^{\dag}a_{k}
             +\sum_{q}\om_{so}b_{q}^{\dag}b_{q} \hspace*{0.2cm}, \\
& &  H'_{e-lo}=u_{1}^{-1}H_{e-lo}u_{1}=\sum_{k}[V_{k}^{\ast}sin(zk_{z})
              a_{k}^{\dag}+h.c.]  \hspace*{0.2cm},\\
& &  H'_{e-so}=u_{1}^{-1}H_{e-so}u_{1}=\sum_{q}[V_{q}^{\ast}exp(-qz)
                exp(-iq\cdot\rho)b_{q}^{\dag}+h.c.] \hspace*{0.2cm},\\
& &  H'_{\perp}=u_{1}^{-1}H_{\perp}u_{1}=H_{\perp}  \hspace*{0.2cm}.
\end{eqnarray}
\hspace*{0.5cm} Following the scheme of Huybrechts, we introduce the
creation and annihilation operators $c_{j}^{\dag}$ and $c_{j}$ by
\begin{eqnarray}
& &          p_{j}=\f{\sqrt{\lam}}{2}(c_{j}^{\dag}+c_{j})\hspace{0.2cm},\\
& &          \rho_{j}=\f{i}{\sqrt{\lam}}(c_{j}-c_{j}^{\dag})\hspace{0.2cm}.
\end{eqnarray}
Where the subscript j refers to the x and y directions, $\lam$ is the
variational parameter,and $c_{j}^{\dag}$($c_{j}$) is Boson operator.\\
\hspace*{0.5cm} Rewriting (9), (11) using (13), (14), one gets:
\begin{eqnarray}
 H'_0=& &\f{\lam}{8}[\sum_{j}(c_{j}^{\dag}c_{j}^{\dag}+c_{j}c_{j})
         +\sum_{j}(c_{j}^{\dag}c_{j}+c_{j}c_{j}^{\dag})]
         -\f{\sqrt{\lam}}{2}\sum_{k,j}a_{k}^{\dag}a_{k}k_{//j}
         (c_{j}^{\dag}+c_{j}) \nonumber\\
 & &+\f{1}{2}\sum_{k}a_{k}^{\dag}a_{k}k_{//}^{2}+\f{1}{2}\sum_{k,k'}
     a_{k}^{\dag}a_{k'}^{\dag}a_{k}a_{k'}k_{//}k'_{//}+
     \f{\om_{c}^2}{8\lam}\sum_{j}(c_{j}c_{j}^{\dag}
      +c_{j}^{\dag}c_{j})  \nonumber\\
& & -\f{\om_{c}^2}{8\lam}\sum_{j}(c_{j}^{\dag}c_{j}^{\dag}+c_{j}c_{j})
    +\f{i\om_{c}}{2}(c_{x}c_{y}^{\dag}-c_{x}^{\dag}c_{y})+
    \sum_{k}\om_{lo}a_{k}^{\dag}a_{k}
    +\sum_{q}\om_{so}b_{q}^{\dag}b_{q} \nonumber\\
 & & +\f{i\om_c}{2\sqrt\lam}\sum_{k}a_{k}^{\dag}a_{k}[k_{x}(c_{y}
    -c_{y}^{\dag})-k_{y}(c_{x}-c_{x}^{\dag})]\hspace{0.2cm},  \\
 H'_{e-so}=& & \sum_{q}[V_{q}^{\ast}exp(-qz)exp(\sum_{j}\f{q_j}{\sqrt{\lam}}
    (c_{j}-c_{j}^{\dag}))b_{q}^{\dag}+h.c.]\hspace{0.2cm},
\end{eqnarray}
Yet, $H'_{e-lo}$ and $H'_{\perp}$ are invariable in forms. Let continue to
do the $u_2$ transformation:
\begin{equation}
   u_2=exp[\sum_{k}(a_{k}^{\dag}f_{k}e^{-ik_{//}\cdot\rho_0}-
   a_{k}f_{k}^{\ast}e^{ik_{//}\cdot\rho_0})+\sum_{q}(b_{q}^{\dag}g_{q}
   e^{-iq\cdot\rho_0}-b_{q}g_{q}^{\ast}e^{iq\cdot\rho_0})]
\end{equation}
Where $f_k$($f_{k}^{\ast}$) and $g_q$($g_{q}^{\ast}$) are variational
parameters, and $\rho_0=(x_0,y_0)$ is the electron orbit center where
$x_0=x/2-p_y/\om_c$, $y_0=y/2+p_x/\om_c$.The necessity of making the
phonon deformation centered at $\rho_0$ was emphasized in an elaborate
discussion by Whitfield {\it et al} [6]. 
\begin{eqnarray}
& & u_2^{-1}a_{k}u_2=a_k+f_{k}e^{-ik_{//}\cdot\rho_0},\hspace*{0.3cm}
    u_2^{-1}a_{k}^{\dag}u_2=a_k^{\dag}+f_{k}e^{ik_{//}\cdot\rho_0},\\
& & u_2^{-1}b_{q}u_2=b_q+g_{q}e^{-iq\cdot\rho_0},\hspace*{0.3cm}
    u_2^{-1}b_{q}^{\dag}u_2=b_q^{\dag}+g_{q}e^{iq\cdot\rho_0}\hspace{0.2cm}.
\end{eqnarray}
\hspace*{0.5cm} Performing $u_2$ transformation and using (18) and
(19), one can get $H''_0$, $H''_{e-lo}$ and $H''_{e-so}$ after lengthy
calculation.\\
(1){\it The ground state:}\\
\hspace*{0.5cm} Determining the ground state $|0>$ in the new representation by
\begin{equation}
        c_j|0>=a_{k}|0>=b_{q}|0>=0     \\
\end{equation}
one gets the energy $E''_{//}(0)$ of the ground state
\begin{eqnarray}
    E''_{//}(0) &=&<H''_{//}>=<H''_{0}+H''_{e-lo}+H''_{e-so}>  \nonumber\\
    &=&\f{\lam}{4}+\f{\om_c^2}{4\lam}+\sum_{k}\om_{lo}|f_k|^2+
      \sum_{q}\om_{so}|g_q|^2+\f{1}{2}\sum_{k}|f_{k}|^2k_{//}^2
      +\f{1}{2}\sum_{k,k'}|f_{k}|^2|f_{k'}|^2k\cdot k'  \nonumber\\
    & &+\sum_{k}[V_{k}^{\ast}sin(zk_z)f_{k}^{\ast}+V_{k}sin(zk_z)f_k]
      exp(-\f{k_{//}^2}{8}(\f{1}{\lam}+\f{\lam}{\om_{c}^2}))\nonumber\\
    & &+\sum_{q}[V_{q}^{\ast}exp(-qz)g_{q}^{\ast}+V_{q}exp(-qz)g_{q}]
       exp(-\f{q^2}{8}(\f{1}{\lam}+\f{\lam}{\om_{c}^2}))\hspace{0.2cm}.
\end{eqnarray}
\hspace*{0.5cm}Minimizing $E''_{//}(0)$ with respect to $f_k(f_{k}^{\ast})$
and $g_q(g_{q}^{\ast})$ and neglecting higher order of $f_{k}$, one gets
\begin{equation}
       E''_{//}(0)=\f{\lam}{4}+\f{\om_{c}^2}{4\lam}+V_{e-lo}(0)+V_{e-so}(0)
\end{equation}
Where $V_{e-lo}(0)$ and $V_{e-so}(0)$ are the correction energy induced by
electron-LO phonon and electron-SO phonon interaction respectively,with
\begin{eqnarray}
& & V_{e-lo}=-\sum_{k}\f{|V_{k}|^{2}sin^2(zk_z)
exp(-\f{k_{//}^2}{4}(\f{1}{\lam}+\f{\lam}{\om_{c}^2}))}
{\f{1}{2}k_{//}^2+\om_lo}\hspace{0.2cm}, \\
& & V_{e-s0}=-\sum_{q}\f{|V_q|^{2}exp(-2qz)exp(-\f{q^2}{4}(\f{1}{\lam}
+\f{\lam}{\om_{c}^2}))}{\om_{so}}\hspace{0.2cm}.
\end{eqnarray}
\hspace*{0.5cm}Considering the explicit forms of $V_k$ and $V_q$, (23) and
(24) can be reduced to
\begin{eqnarray}
& & V_{e-lo}=-\f{\sqrt{2}\alpha_l\om_{lo}^\f{1}{2}}{2}
    \int_0^{\infty}\f{(1-e^{-2k_{//}z)}exp(-\f{k_{//}^2}{4}
    (\f{1}{\lam}+\f{\lam}{\om_{c}^2}))}{1+\f{k_{//}^2}{2\om_{lo}}}dk_{//}\hspace{0.2cm},\\
& & V_{e-so}=-\sqrt{2}\alpha_{s}\om_{so}^{\f{1}{2}}\int_{0}^{\infty}
    exp^{(-2qz)}exp(-\f{q^2}{4}(\f{1}{\lam}+\f{\lam}{\om_{c}^2}))dq\hspace{0.2cm}.
\end{eqnarray}
\hspace*{0.5cm} In eq.(22) ,if neglecting $V_{e-lo}(0)$,$V_{e-so}(0)$ and
minimizing $E''_{//}(0)$ with respect to $\lam$, one can obtain
$\lam=\om_c$ and $E''_{//}(0)$=$\f{1}{2}\om_c$. Obviously, this is the
electron ground state energy of Landau Level.\\
\hspace*{0.5cm}The effective Hamiltonian of the polaron system in the
ground state $H_{eff}$ is
\begin{equation}
        H_{eff}=E''_{//}(0)+H_{\perp}=\f{p_{z}^2}{2}+\f{\lam}{4}+
              \f{\om_c^2}{4\lam}+V_{e-lo}+V_{e-so}+V_{img}\hspace{0.2cm},
\end{equation}
with
$$
 V_{img}=\f{e^2(\var_{\infty}-1)}{4z(\var_{\infty}+1)}\hspace{0.2cm}.
$$
(2){\it The excited state:}\\
\hspace*{0.5cm} The excited state $|1>$ in the new representation is 
\begin{equation}
        |1>=c_{j}^{\dag}|0>=a_{k}^{\dag}|0>=b_{q}^{\dag}|0>     
\end{equation}
considering
 \begin{equation}
 <1|exp(-iq\cdot\rho_0)|1>=exp(-\f{q^2}{8\lam}-\f{{\lam}q^2}{8\om_{c}^2})
                     [1-(\f{q^2}{8\lam}+\f{{\lam}q^2}{8\om_{c}^2})]
\end{equation}
similarly, one gets
\begin{equation}
       E''_{//}(1)=\f{3\lam}{4}+\f{3\om_{c}^2}{4\lam}+\om_{lo}+\om_{so}+V_{e-lo}(1)+V_{e-so}(1)
\end{equation}
Where $V_{e-lo}(1)$ and $V_{e-so}(1)$ are the correction energy induced by
electron-LO phonon and electron-SO phonon interaction of the excited state respectively,with 
\begin{eqnarray}
& & V_{e-lo}(1)=-\sum_{k}\f{|V_{k}|^{2}sin^2(zk_z)
exp(-\f{k_{//}^2}{4}(\f{1}{\lam}+\f{\lam}{\om_{c}^2}))
[1-(\f{k_{//}^2}{8\lam}+\f{\lam k_{//}^2}{8\om_{c}^2})]^2}
{\f{1}{2}k_{//}^2+\om_lo}\hspace{0.2cm}, \\
& & V_{e-s0}(1)=-\sum_{q}\f{|V_q|^{2}exp(-2qz)exp(-\f{q^2}{4}(\f{1}{\lam}
+\f{\lam}{\om_{c}^2}))[1-(\f{q^2}{8\lam}+\f{\lam q^2}{8\om_{c}^2})]^2}{\om_{so}}\hspace{0.2cm}.
\end{eqnarray}
{\bf 3. Discussion and Conclusion}\\
\hspace*{0.5cm} Taking the polaron in  polar crystal AgBr as an example,
we perform a numerical evaluation. The parameters concerned are listed as
follows: $\var_0$=10.6, $\var_{\infty}$=4.68, $\hbar\om_{lo}$=17.1meV,
$\hbar\om_{so}$=14.5meV, $\alpha_l$=1.56, $\alpha_s$=2.56, $m_b$=0.22m,
m is the free electron mass. All the above parameters are taken from [12].
\begin{center}
      Fig.1    \hspace*{3cm}                        Fig.2
\end{center}
\hspace*{0.5cm}Fig.1 shows the relationships between the induced potential
$|V_{e-lo}(0)|$ resulting from the electron-LO phonon interaction, the induced
potential $|V_{e-so}(0)|$ resulting from electron-SO phonon interaction, and
their sum $|V_{e-lo}(0)|+|V_{e-so}(0)|$ and the coordinate z. From fig.1 we can
see that the induced potential $V_{e-lo}(0)$ increase with increasing
coordinate z, whereas $V_{e-so}(0)$ decrease with increasing coordinate z.
Compared with the weak electron-phonon interaction case [13], $|V_{e-so}(0)|$
and $|V_{e-lo}(0)|$ are larger than the corresponding weak coupling ones,yet
their lines tendency is similar to the weak coupling ones.
When $z<10\AA$, $|V_{e-so}(0)|$ is larger than $|V_{e-lo}(0)|$,
whereas for $z>35\AA$, $|V_{e-so}(0)|$ becomes smaller compared with
$|V_{e-lo}(0)|$ and so $|V_{e-lo}(0)|$ is the dominant term.
$|V_{e-lo}(0)|+|V_{e-so}(0)|$ decreases with increasing z. When $z>35\AA$,
$|V_{e-lo}(0)|+|V_{e-so}(0)|$ approaches $|V_{e-lo}(0)|$ because $|V_{e-so}(0)|$ is
much smaller compared with $|V_{e-lo}(0)|$.\\
\hspace*{0.5cm}Fig.2 show the induced potentials as a function of the 
electron coordinate z for different magnetic fields. In fig.2, the
dot-line stands for the case when B=4T and the solid-line stands for
the B=8T case. From fig.2, we can see that $|V_{e-lo}(0)|$ and $|V_{e-so}(0)|$
for B=8T are larger than the corresponding ones for B=4T. This result
show that external magnetic field can enhance the interaction between
an electron and phonons. Let us give a possible physical explanation of
the result. When magnetic field is applied, the polarizability of the
crystal increases and the density of phonons increases. Because the
strength of the electron-phonon interaction is proportional to the
number of phonons, the interaction energies, for both $|V_{e-lo}(0)|$ and
$|V_{e-so}(0)|$, increase with increasing magnetic field strength.
\begin{center}
      Fig.3   
\end{center}
\hspace*{0.5cm}Fig.3 illustrates the energy difference induced by the electron-LO (SO) phonon
interaction between the excited and ground states . From fig.3 one can see that the difference between
$V_{e-so}(1)$ and $V_{e-so}(0)$ decreases , yet the difference between 
$V_{e-lo}(1)$ and $V_{e-lo}(0)$ increases as the coordinate z increases.
As z is small ($z<10\AA$) the magnitude of $V_{e-so}(1)-V_{e-so}(0)$ is much large than that of 
$V_{e-lo}(1)-V_{e-lo}(0)$, and when $ z>33\AA$ $|V_{e-so}(1)-V_{e-so}(0)|$ becomes less than 
$|V_{e-lo}(1)-V_{e-lo}(0)|$. These results give further implication that the electron-SO phonon interaction 
play a dominant role as electron are near the polar crystal surface. \\
\hspace*{0.5cm}In summary, by applying variational method, the strong electron-surface
optical(SO) phonon interaction and the weak electron-longitudinal optical
(LO) phonon interaction are studied systematically in present paper. The
formula of induced energies $V_{e-so}(0)$, $V_{e-lo}(0)$,$V_{e-so}(1)$,$V_{e-lo}(1)$
are given. The numerical results show that as the distance between the
electron and the polar crystal surface increases, the electron-LO phonon
interaction energy increases yet the electron-SO phonon interaction energy
decreases. The numerical results also show that both electron-LO and
electron-SO phonon interaction are enhanced as the magnetic field
strength increases.\\
\newpage
\baselineskip 0.7cm
{\large References}\\
{ [1] S. D. Sarma,  and  A.Madhukar, {\it Phys. Rev.} {\bf B22}(1980)2823}\\
{ [2] D. V.Melnikov  and W.B.Fowler, {\it Phys. Rev.} {\bf B63} (2001)1653020}\\
{ [3] W. J. Huybrechts, {\it J. Phys. C: Solid state phys.} {\bf 10}(1977)3761}\\
{ [4] N. J. Tokuda,  {\it J. Phys. C: Solid state phys.} {\bf 13}(1980)851}\\
{ [5] Sh.W. Gu,  Y. C. Li, and  L. F. Zheng, {\it Phys. Rev.} {\bf B39}(1989)1246}\\
{ [6] G. Whitfield,  R. Parker, and  M. Rona,{\it Phys. Rev.} {\bf B13}(1976)2132}\\
{ [7] A. Ercelebi, {\it J. Phys.: Condens. Matter} {\bf 1}(1989)2321}\\
{ [8] G. Q. Hai, F. M. Peeters, and J. T. Devreese,  {\it Phys. Rev.} {\bf B47}(1993)10358}\\
{ [9] Y. J. Wang, H. A. Nickle,and B. D. McCombe,{\it et al,} {\it Phys. Rev. Lett.} {\bf 79}(1997)3226}\\
{[10] X.K.Hua, Y.Z.Wu and Zh.Y.Li {\it Chin.Phys.} { Vol.\bf12}(2003)No.11 635}\\ 
{[11] J. S. Pan,  {\it Phys. Status Solidi} {\bf b128}(1985)307}\\
{[12] E. Kartheuser, {\it Polarons in Ionic crystals and polar
      semiconductors}, North-Holland, \\ \hspace*{0.5cm} Amsterdam(1972)718 }\\
{[13] B. H. Wei, K. W. Yu, and  F. Ou, {\it J. Phys.: Condens. Matter} {\bf 6}(1994)1893}
\newpage
\begin{center}
{\bf Figure Captions}
\end{center}
Fig.1.  $|V_{e-so}(0)|$, $|V_{e-lo}(0)|$, and $|V_{e-so}(0)|+|V_{e-lo}(0)|$
as a function of coordinate z.\\
\\
Fig.2.  $|V_{e-so}(0)|$ and $|V_{e-lo}(0)|$ as a function of coordinate z for
different magnetic fields. The solid line and the dot line represent B=8T
 and B=4T, respectively.\\

Fig.3 $|V_{e-ph}(1)-V_{e-ph}(0)|$ as a function of coordinate z at B=10T.\\
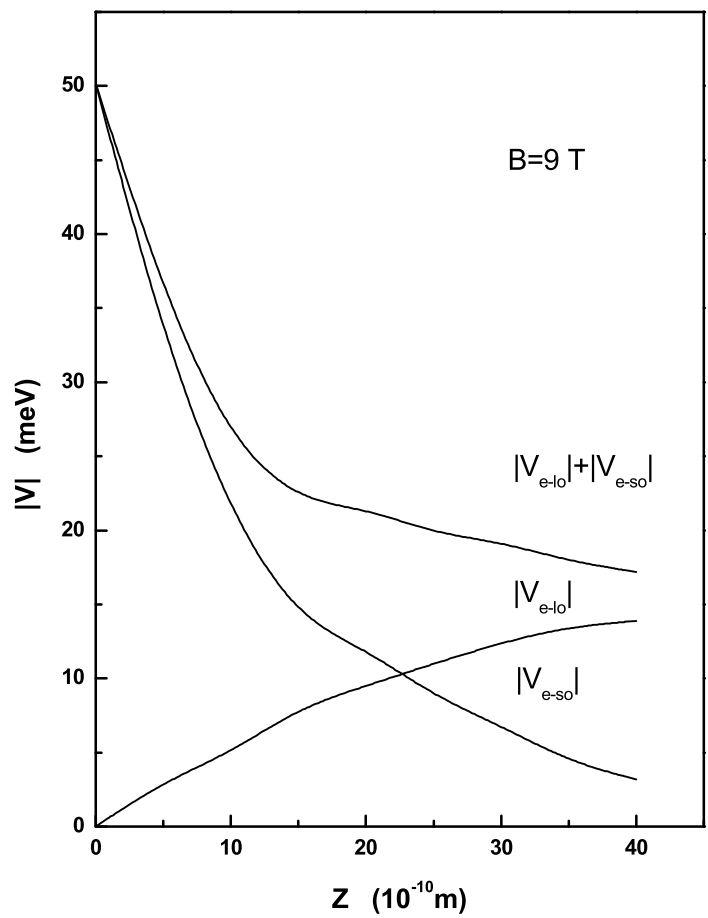
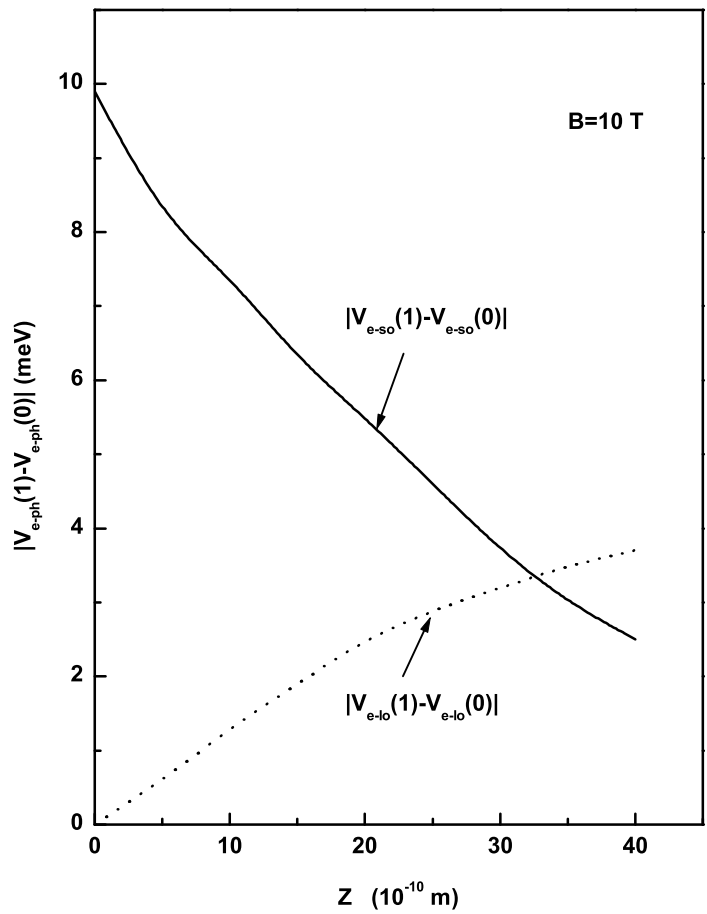
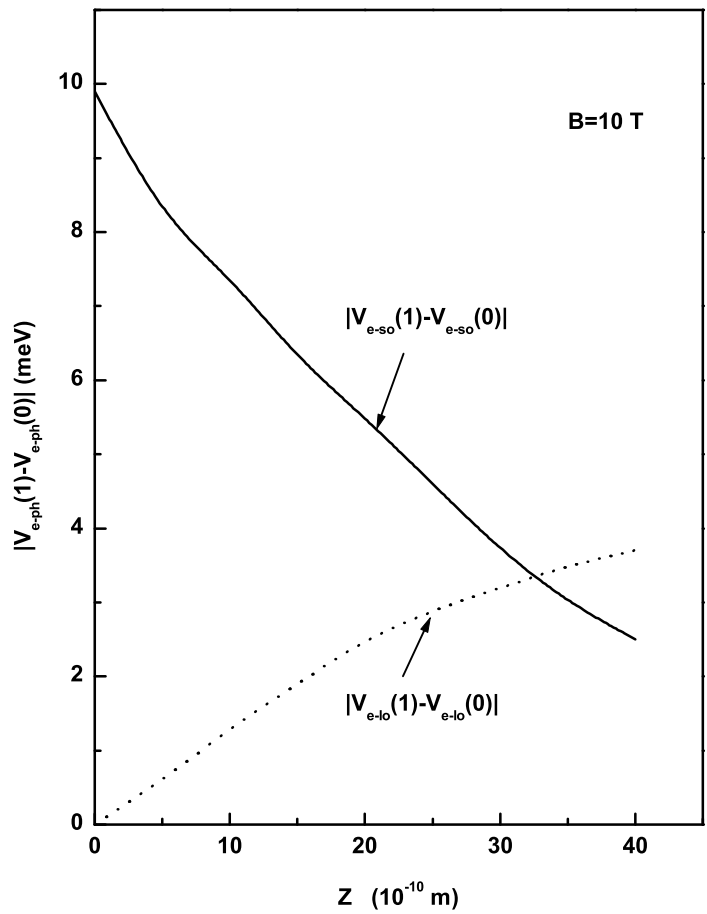
\end{document}